\documentstyle[11pt]{article}
\begin{document}
\topmargin 0cm
\baselineskip=.7cm
\parskip=.2cm
\textwidth=14cm
\textheight=21cm
\def\theequation{\arabic{section}.\arabic{equation}}
\newcommand{\beq}{\begin{equation}}
\newcommand{\eeq}{\end{equation}}
\newcommand{\rbracket}{\right]}
\newcommand{\fff}{{\bar f}}
\newcommand{\lbracket}{\left[}
\newcommand{\DD}{{\cal D}}
\newcommand{\DDt}{{\nabla}_{t}}
\newcommand{\ddt}{\partial_t}
\newcommand{\ddx}{\partial_x}
\newcommand{\h}{\theta}
\newcommand{\Ag}{A^{g}}
\newcommand{\aaa}{{\alpha}}
\newcommand{\Q}{\bar{Q}}
\newcommand{\Pb}{\bar{P}}
\newcommand{\ppb}{\bar p_{i}}
\newcommand{\ps}{\bar{\psi}}
\newcommand{\LL}{{\cal L}}
\newcommand{\ggg}{\sf g}
\newcommand{\curve}{\Sigma_{\tau}}
\newcommand{\intc}{\int_{\Sigma_{\tau}}}
\newcommand{\haha}{\theta_{11}}
\newcommand{\hgg}{\hat g}
\newcommand{\elg}{{\hat {\sf g}}^{\Sigma_{\tau}}}
\newcommand{\eld}{({\hat {\sf g}}^{\Sigma_{\tau}})^{*}}
\newcommand{\Rn}{{\cal R}_{\nu}}
\newcommand{\ppz}{\phi (z, {\bar z})}
\newcommand{\aaz}{{\bar A} (z, {\bar z})}
\newcommand{\ggz}{g (z, {\bar z}) }
\newcommand{\slnc}{{\it sl}_{N} ({\bf C})}
\newcommand{\SLnc}{{\it SL}_{N} ({\bf C})}
\newcommand{\sun}{{\it su}_{N}}
\newcommand{\ep}{\epsilon_{i}}
\newcommand{\Vq}{\Delta (\Q)}
\newcommand{\Vp}{\Delta (\Pb)}
\newcommand{\RR}{{\cal R}_{\nu_{i}}}
\newcommand{\rw}{\rightarrow}
\def\op{operator}
\def\tly{topologically}
\def\onlabs{on leave of absence from}
\def\itep{ Institute of Theoretical and Experimental Physics}
\def\exl{external}
\def\CG{Clebsh-Gordon}
\def\gl{global}
\def\f{field}
\def\exst{existence}
\baselineskip = 3mm
\def\Clgr{Calogero}
\def\gl{global}
\def\an{anomaly}
\def\hh {\hat h}
\def\exst{existence}
\def\nl{normalizable}
\def\nty{normalizability}
\def\cn{condition}
\def\con{configuration}
\def\beq { \begin{equation}}
\def\eeq { \end{equation}}
\def\lw {\leftarrow}
\def\tt {\cal t}
\def\intt {\int_{S^{1}}}
\def\bbb {\box}
\def\lll {\lambda}
\def\ddd {\cal D}
\def\con{configuration}
\setlength{\oddsidemargin}{-0.1in}
\title{ Integrable many body systems in the field theories }
\author{\sf Alexander Gorsky
\thanks{permanent address:ITEP,117259,B.Cheremushkinskaya,
Moscow,Russia }\\
\centerline{\em  Institute of Theoretical Physics at Uppsala
University }\\
\centerline {Talk given at the Alushta Conference,June 1994}}

\date{
\setlength{\unitlength}{\baselineskip}
\begin{picture}(0,0)(0,0)
\put(12,15){\makebox(0,0){UUITP-16/94}}
\put(12,14.4){\makebox(0,0){October 1994}}
\put(12,13.7){\makebox(0,0){hepth/xxxddd}}
\end{picture}}

\maketitle
\abstract{We review recent results which clarify the role of the
integrable many-body problems in the quantum field theory
framework.They describe the dynamics of
the topological degrees of freedom in the theories which are obtained
by perturbing the topological ones by the proper Hamiltonians and
sources.
Interpretation of the many-body dynamics as a motion on the different
moduli spaces as well as the property of duality is discussed.Tower
of many-body systems can be derived from a tower of the field
theories with appropriate phase spaces which have a transparent
interpretation in terms of the group theory.The appearance of
Calogero-type systems in different physical phenomena is mentioned.}



\newpage
\setcounter{footnote}0

\section{Introduction}

Recently the integrable many body systems have appeared on the scene
after the long delay in a new image.Fifteen years ago they were
considered as a simplest example of integrable systems which have
the nice interpretation in terms of the group theory.The most popular
examples of these systems are famous Calogero-Sutherland-Moser(CSM)
and Toda ones.\cite {moser}  \cite{toda}.They were discovered
heuristically and later the transparent interpretation was
given.Excellent review of the activity on this subject in seventies
and the beginning of eighties can be found in \cite {olsh-per-cl}
\cite {per-ol-q}.The main outcome of the investigations was the
understanding that for each root system the corresponding many body
problem can be obtained in a canonical way. Moreover the wave
functions and the spectrum allow the group theory derivation.For
example  characters and zonal spherical functions appear to be the
wave functions for  the particular coupling constants.Another
important observation was that the many-body Hamiltonians are closely
related with the radial parts of the Laplace operators on the
symmetric spaces.

 Among further achievements in this direction it is worth mentioned
two important results.At first it was realized that these systems
have very transparent phase spaces which can be derived via the
general procedure

of the Hamiltonian reduction.
\cite{kag} \cite{Olper2}.Secondly it was discovered by Ruijsenaars
\cite{ruu} that there exists a natural generalization of the systems
which gives rise to the relativistic invariant Hamiltonians closely
related with the quantum groups.Just these facts  occurred to be of
the most importance .The first one gives the correct framework for
the path integral treatment of the problem while the second one
provides with a correct guess for the correspondence between the
tower of many-body Hamiltonians and the tower of topological field
theories.

It was recognized last years that CSM systems appear naturally in
field theoretic models like 2D Yang-Mills theory or gauged WZW one.
The corresponding field theories should have the finite number
degrees of freedom which describe the zero mode variables which then
appear to be the coordinates on the many-body problem phase
space.Thus  the

quantum mechanical phase space is present in the problem from the
very beginning.Then one has to add a Hamiltonian which results in
the transition from the topological

to the physical theory,as it can be easily seen in YM case,and insert
some observable.In the 2D cases Wilson line in an appropriate
representation is inserted.The resulting dynamics on zero modes
coincides with the one for the corresponding many-body problems.In
this way
the relation between 2D YM with inserted Wilson line with
trigonometric

CSM as well as G/G WZW and Ruijsenaars trigonometric system
was established \cite{gor2d,gorrui}.Path integral representation for

the field theories provides with the wave functions and spectrum for
its quantum mechanical counterpart.

A more general look on the problem is also instructive.It can be
formulated in the following manner.Any topological theory can be
reduced to the appropriate moduli spaces like moduli space of the
flat connections or  holomorphic G bundles.Then the moduli space
itself or its cotangent bundles play the role of the phase spaces for
the QM systems so the dynamics proceeds on the particular moduli
space.In all cases a kind of the Wilson line is present so the moduli
spaces for the surfaces with the marked points are of the
interest.The situation is most transparent in the elliptic Calogero
system \cite {gnell} when the phase space is nothing but the
cotangent bundle to the moduli space of the holomorphic G bundle on a
torus with a marked point.Thus it is an example of Hitchin integrable
system on the moduli space \cite {hitchin} and the path integral
representation for elliptic Calogero system gives a reasonable
approach for its quantization.Therefore the problem reduces to the
integrable motion on the moduli spaces so the information obtained
from the quantum many-body problems is useful for the investigation
of the structure of the moduli space and vise versa.

Besides the investigation of the structure of the many-body systems
itself
which is interesting by its own these systems appear in the

different physical situations.
For example hyperbolic Ruijsenaars system describes multisolitonic
sectors in the sin-Gordon theory \cite {solitons} and trigonometric
Calogero one is related with the edge excitations dynamics in the

Quantum Hall Effect \cite{QHE} as well as the anyonic statistics
\cite{anyons}.This shouldn,t be a surprise of course because of the
following reasons.It is known for a long time \cite{manton} that the
geodesic motion on the moduli spaces corresponds to the scattering of
the solitonic like objects in the nonlinear field theory relevant for
this particular moduli space.Just this happens in the sine-Gordon
case which allows the zero-curvature condition formulation,so it is
related with the moduli space of flat connection which is the phase
space for the trigonometric Ruijsenaars system.From the other hand in
the appropriate situation in some complicated case like in QHE one
can look for the topological sectors of the theory and find himself
with the quantum mechanical integrable many-body systems describing
these degrees of freedom.

One more problem,where the Calogero type model appears is c=1 matrix
model.It was shown \cite{avan} that large N limit of  the Calogero
system

results in c=1 collective field theory.Therefore the Calogero model

itself can be considered as its particular discretization.

In this review we try to explain the main features of the picture
,consider the examples available and formulate the problems to be
solved.

\section{CMS system in 2D YM theory}
Let us formulate the approach to the CMS systems based on the
Hamiltonian reduction.This approach occurs to be useful one when
performing the generalization finite-dimensional situation to the
field theory. Classical CSM system is a system of N particles on the
real line
with the pair-wise interaction potential \cite{moser}:
\beq \label{pot}
V(x_{1},\ldots,x_{N}) = \sum_{i\neq j}
\frac{g^{2}}{(x_{i}-x_{j})^{2}},
\eeq
$g^{2}$ is a coupling constant, which is supposed to satisfy $g^{2} >
- 1/8$,
to avoid the collapse of the system.
It is well known \cite{kag} that such a system (and all
generalizations, like Sutherland's one with $\frac{1}{sin^2
(x_{i}-x_{j})}$
instead of $\frac{1}{(x_{i}-x_{j})^{2}}$ or with extra quadratic
potential)
appears as a result of hamiltonian reduction of some simple
hamiltonian
system. For pure Calogero system  it goes as follows.
We start from the free system on the cotangent bundle to the Lie
algebra
$su(N)$ .This phase space is the simplest one in the family related
with the many-body problems.  We consider the space of pairs $(P,Q)$,
where $P$
and $Q$ are $su(N)$ matrices  with
canonical symplectic structure
\beq \label{sym1}
\Omega = tr(\delta P \wedge \delta Q).
\eeq
Free motion is generated by quadratic hamiltonian
\beq \label{h2}
H_{2} = \frac{1}{2} tr P^2
\eeq
On the  $T^{*}su(N)$  acts unitary group $SU(N)$ by adjoint action on
$Q$'s
and coadjoint on $P$'s. This action preserves  symplectic structure.
If we
identify the cotangent bundle with two copies of the Lie algebra with
help of
the Killing form we would get a moment map of this action in the form
\beq \label{mom1}
\mu = [P,Q]
\eeq
According to the standard prescription of hamiltonian reduction
procedure we
restrict ourselves to some level $\mu$ of momentum map and then
factorize
along the distribution of the kernels of the restriction of
symplectic form.
Generally these ones are spanned by the vectors, tangent to the
orbits of the
action of the stabilizer of point $\mu$ in the coadjoint
representation.
In our situation stabilizer depends on the number of different
eigenvalues of
matrix $\mu$.In particular, if matrix $\mu$ has only two different
eigenvalues, with multiplicities $1$ and $N-1$, then stabilizer will
be
$S(U(1) \times U(N-1))$.
We shall denote it by $G_{\mu}$. Now we should resolve equation
(\ref{mom1})
modulo transformations from this group. It is easy to show, that if
matrices
$P$ and $Q$ satisfy condition $[P,Q]=\mu$ then we can diagonalize $Q$
by
conjugation by matrices from the $G_{\mu}$. Let $q_{i}$ be its
eigenvalues.
Then $P$ turns out to be of the form (in the basis,where $Q$ is
diagonal and
$\mu_{ij}=\nu(1-\delta_{ij})$)

\beq \label{ppp}
P=diag(p_{1},\ldots,p_{N})+ \nu
||\frac{1-\delta_{ij}}{(q_{i}-q_{j})}||
\eeq

where $\nu (N-1)$ and $-\nu$ are the eigenvalues of $\mu$.

Symplectic structure on the reduced manifold turns out to be the
standard one,
i.e. $p$'s and $q$'s are canonically conjugate variables.
Hamiltonian (\ref{h2}) is now hamiltonian for the natural system
with potential $V$ (\ref{pot}). In fact, it is possible to obtain
integrable system, starting from the
Hamiltonian
\beq \label{hk}
H_{k} = \frac{1}{k} tr P^k
\eeq
All $H_{k}$ commute among themselves with respect to Poisson
structure,
defined by (\ref{sym1}), thus yielding the complete integrability of
the
classical Calogero model. The restriction to the $su(N)$ case imply
just the
fixing of the center of mass of the system at the point zero.

Now we proceed to the quantization of the Calogero system. We
consider path
integral representation for the wave functions and by means of exact
evaluation of path integral we  obtain finite-dimensional integral
formula for Calogero wave function (see \cite{per-ol-q}).
We shall consider integral over the following set of fields.
First, there will be the maps from the time interval $[0,T]$ to the
direct
product of three copies of $su(N)$ Lie algebra, actually to the
product $T^{*}{su(N)}\times {su(N)}$. We denote corresponding
matrix-valued functions as $P(t)$,$Q(t)$ and $A(t)$
respectively. Then, we have fields $f(t)$ and $f^{+}(t)$ which are
the $CP^{N-1}$ - valued fields. Locally, these are $N$ complex
numbers
$f_{i}$, which satisfy the condition $\sum_{i=1}^{N} f_{i}^{+} f_{i}
= 1$ and are considered up to the multiplication $f \rw e^{i \h} f$,
$f^{+} \rw f^{+} e^{-i \h}$.

Now let us define the action. We shall do it in two steps. At first
we define
the action for $P$, $Q$, $A$ fields. It will be a sum of action of
unreduced free system plus term which will fix the value of the
moment map. Field $A$ will play the role of the lagrangian
multiplier.

\beq \label{act1}
S_{P,Q,A} = \int tr ( i P{\ddt}Q - \frac{1}{2} P^2 + i A ([ P,Q ]  +
i \mu))
\eeq

Since $\mu$ has only two different eigenvalues with multiplicities
$1$ and
$N-1$, it has the following form:

\beq \label{mu}
\mu = i \nu ( Id - e \otimes e^{+} )
\eeq

where  $\nu$ is arbitrary (up to now) real number and we can choose
vector $e$
to be $e = \sum_{i=1}^{N} e_{i}$, $e_{i}$ being the standard basis in
$C^{N}$.

We shall evaluate the transition amplitude $<q'|\exp((t'-t)H)| q>$.
It is given by the path integral with boundary conditions.
First we integrate out the $A$ field. We observe that our path
integral is
invariant under the (small) gauge transformations
\begin{eqnarray}
&&Q \rw g Q g^{-1} \; , \; P \rw g P g^{-1}
\nonumber\\
&&A {\rw} {\Ag} = g A g^{-1} + i({\ddt}g) g^{-1};
\nonumber \\
&&g(t) \in G_{\mu} \; , \; g(0) = g(T) = Id \label{gauge1}
\end {eqnarray}

We can enlarge the symmetry of our integral by introducing auxilliary
fields
$f^{+},f$ as follows. Let us rewrite term $tr(\mu (\Ag))$ in terms
of the vector field ${\fff} = g^{-1} e$, with $e$ from the previous
chapter. We get
\beq \label{cpn}
S_{f} = \nu \int_{0}^{T} {\fff}^{+} (i \ddt - A ) {\fff}
\eeq

Obviously, $S_{f}$ depends only on the class of $\fff$'s  modulo
the total factor $e^{i \h}$ if at the ends $0$ and $T$ $\h$'s
coincide.
Hence, it depends only on the $CP^{N-1}$ valued field $f$, which
corresponds to the $\fff$. Therefore,  our action $S = S_{P,Q,A} +
S_{f,f^{+}}$ acquires the form
\beq \label{actcpn}
S = tr ( i P {\DDt} Q - \frac{1}{2}P^{2} + i \nu f^{+} {\DDt} f ),
\eeq
where ${\DDt} Q = {\ddt} Q + i[ A, Q ]$, ${\DDt} f = {\ddt} f + i A
f$.
It is clear, that this action has $SU(N)$ gauge symmetry:
\begin{eqnarray}
&& P \rw P^{g} = g P g^{-1}, \;  Q {\rw} Q^{g} = g Q g^{-1}
\nonumber\\
&&A {\rw} {\Ag} = g A g^{-1} + i({\ddt}g) g^{-1}
\nonumber \\
&&f \rw g f \; , f^{+} \rw f^{+} g^{-1} \; , g \in SU(N)
\label{gauge2}
\end {eqnarray}
This invariance holds in the sense that amplitude is invariant under
the
$SU(N)$ action:
$<Q,f|\ldots|Q',f'> = <g Q g^{-1}, g f|\ldots| g' Q' g'^{-1}, g'
f'>$.

As it usually goes in matrix models we diagonalize matrix $Q(t)$ by
the
conjugation by unitary matrix $g(t)$. We denote diagonalized matrix
$Q(t)$ by
$\Q$, and let $q_{i}$ be its entries.  Due to $SU(N)$ invariance of
the action
"angle" variables $g(t)$ decouple except the boundary terms.
It is a gauge fixing so Faddeev-Popov determinant arises, which is
nothing
but the square of the Vandermonde determinant of $q_{i}$ , ${\Vq}^{2}
=
\prod_{i<j}(q_{i}-q_{j})^{2}$.

We can keep it just by  decomposing measure on the Lie algebra as the
product
of the Cartan measure on the $q_{i}$ times ${\Vq}^{2}$ times the Haar
measure
on the unitary group and it should be done in each point of time
interval, i.e. ${\DD}Q = \prod_{t} d{\Q} {\Delta({\Q}_{t})}^{2} dg$.
Actually we get here an extra integration over Cartan subgroup. This
implies
that our gauge fixing is incomplete. We approve it by fixing the $f$
field
to be real and positive. . This can be always achieved in the generic
point due to
the remainig action of the Cartan subgroup $T^{N-1}$.

We will need the fact that after a quantization of $CP^{N-1}$ with
symplectic structure $\Omega$ some irreducible representation
$R_{\nu}$ of $SU(N)$ appears.
with dimension
$$
dim R_{\nu} = \frac {(N + \nu -1)!}{\nu! (N-1)!}
$$
This information permits  us  partially  integrate  out  fields  in
our
theory, namely, we integrate $f^{+},f$ and this make us left with the
matix element of the ${\cal  P}  \exp  \int  A$  in  the
representation
$R_{\nu}$.
On the other hand, if we look at the integral over the non-diagonal
part
of the $A$ field,  we  just  recover  constraint  (\ref{mom1})  and
the
Vandermonde - Faddeev - Popov determinant cancels.
Here we don't introduce ghost fields, but in principle it can be done
and
the result will be the same.
We have gauged $Q$ to be in the Cartan subalgebra, therefore
the term $[P, Q]$ has become orthogonal to the Cartan part of the
$A$.
So the integral over $CP^{N-1}$ valued part of the fields  is  done
and
the remaining integral over $P,Q,A$ recovers Calogero amplitude for
$g^{2} = \nu (\nu -1)$.

The generalization to the case of affine group is straightforward and
is presented below.It occurs that the generalization of action for
Calogero system is nothing but 2D YM action  on the cylinder with
inserted Wilson line in the representation which defines the coupling
constant for the CSM system.

We start from consideration of the central extension of
current algebra ${\LL}{\ggg}$, where $\ggg$ is some semi-simple Lie
algebra.
The cotangent bundle to this algebra consists of the sets
$(A,k;\phi,c)$,
where $\phi$ denotes $\ggg$-valued scalar field on the circle, $c$ is
a central element, $A$ is $\ggg$-valued one-form on the circle and
$k$
is the level, dual to $c$.
$$
[({\phi}_{1}({\varphi}),c_{1}), ({\phi}_{2}({\varphi}),c_{2})] =
([{\phi}_{1}({\varphi}),{\phi}_{2}({\varphi})],
\int_{S^{1}} <{\phi}_{1}, {\partial}_{\varphi} {\phi}_{2}>),
$$

Now let us turn to the dual space to the $\hat g$. This space ${\hat
g}^{*}$
consists
of the pairs $(A, {\kappa})$
where $A$ is $g$-valued one-form (actually, a distribution) on
the circle and
$\kappa$
is just the real number. The pairing between the $\hat g$ and
${\hat g}^{*}$ is

$$
<(A, {\kappa}) ; (\phi , c)> = \intt <\phi, A> + c {\kappa}.
$$

The direct sum ${\hat g} \oplus {\hat g}^{*}$ we will denote as
$T^{*}{\hat g}$.
On this cotangent bundle the natural symplectic structure is defined:

\beq
\Omega = \intt {\sf tr} ( \delta \phi \wedge \delta A) + \delta c
\wedge \delta {\kappa}
\label{sympl}
\eeq

We can define adjoint and coadjoint actions of
the loop group ${\cal L}G$
on the $\hat g$ and ${\hat g}^{*}$. The former is defined from the
commutation
relations and the latter from the pairing, that is the element
$g(\varphi)$
acts as follows:

\beq
(\phi(\varphi) , c) \rw (g(\varphi) \phi(\varphi) g(\varphi)^{-1},
\intt {\sf tr} (- \phi g^{-1}{\partial_{\varphi}}g) + c )
\label{adj}
\eeq

\beq
(A, {\kappa}) \rw (g A g^{-1} + {\kappa} g {\partial_{\varphi}}
g^{-1},
{\kappa})
\label{coadj}
\eeq

This action clearly preserves the symplectic structure (\ref{sympl})
and thus
defines a moment map

$$
\mu : T^{*}{\hat g} \rw {\hat g}^{*}
$$

which sends a 4-tuple $(\phi, c; A, {\kappa})$ to the pair
$( {\kappa} d \phi + [A, \phi] , 0)$.

 Now let us choose an appropriate level of moment map and make a
hamiltonian
reduction under this level. To guess, what element of ${\hat g}^{*}$
we should
choose it is convenient to return back to the finite-dimensional
example
of this
procedure. Namely, if one considers a reduction of the $T^{*}g$
with respect to
the action of $G$ by conjugation, then, to get a desired Calogero
system,
one takes
an element $J$ of $g^{*}$, which has maximal stabilizer, different
from
the whole $G$.
It is easy to show, that the representative of the coadjoint orbit of
this
element has the following form:

\beq
J_{\nu} = {\nu} \sum_{\alpha \in \Delta_{+}}( e_{\alpha} +
e_{-\alpha}),
\label{mom}
\eeq

where $\nu$ is some real number,  $e_{\pm \alpha}$ are the elements
of
nilpotent subalgebras $n_{\pm} \subset g$, which correspond to the
roots
 $\alpha$, and ${\Delta}_{+}$ is the set of positive roots. Let us
denote the coadjoint orbit of $J_{\nu}$ by ${\cal O}_{\nu}$
and by the $R_{\nu}$
the representation of $G$, arising upon the quantization of ${\cal
O}_{\nu}$.
For generic $J \in g^{*}$ let
$G_{J}$ will denote the stabilizer of $J$ in the coadjoint orbit
${\cal O}_{J}$
of $J$, i.e. ${\cal O}_{J} = G / G_{J}$.

 Now let us return back to the case of affine Lie algebra.
  An appropriate value of
 the moment map we are
looking for is the following:

\beq
\mu = ({\cal J}[{\mu}],0) : {\cal J}[{\mu}]({\varphi}) =
\delta({\varphi})
J_{\nu}
\label{affmom}
\eeq

The coadjoint orbit of $\mu$ is nothing, but the finite-dimensional
orbit
${\cal O}_{\nu}$.

 Now let us complete the reduction. To this end we should resolve the
equation,which plays the role of the Gauss law constraint in the
gauge theory

\beq
{\kappa} \partial_{\varphi} \phi + [A, \phi] =  J_{\mu}({\varphi}) =
\delta({\varphi}) J_{\nu}
\label{redeq}
\eeq

modulo the action of stabilizer of $\mu$, that is modulo subgroup of
${\cal L}G$,
consisting of $g({\varphi}) \in G$, such that $g(0) \in G_{\nu}$. We
can do
it as follows.
First we use generic gauge transformation $\tilde g({\varphi})$ to
make $A$
to be the Cartan subalgebra $t \subset g$
-valued constant one-form $D$ (it is always possible for
non-vanishing
$\kappa$). We are left with the
freedom to use the constant gauge transformations with values in
the Cartan subgroup
${\bf T} \subset G$, generated by $D$ - these do not touch $D$.
Actually, the choice of $D$ is not unique. The only
invariant of ${\cal L}G$ action is  the conjugacy class of monodromy
$\exp ( \frac{2\pi}{\kappa} D) \in {\bf T} \subset G$. Let us fix
some of
these choices. Let
${\sf i} x_{i}$ will denote the entries of $D = iX$. Let us decompose
the $g$-valued function
$\phi$ on the $S^{1}$ on its Cartan-valued part $P(\varphi) \in t$
and let ${\phi}_{\pm}({\varphi}) \in n_{\pm}$ be its
nilpotent-valued parts. Let ${\phi}_{\alpha} = <{\phi}, e_{\alpha}>$.
 Then the equation (\ref{redeq}) will take the form:

\beq
{\kappa}\partial_{\varphi} P = \delta({\varphi}) [ J_{\nu}^{g}
]_{\gamma}
\label{careq}
\eeq

\beq
{\kappa}\partial_{\varphi}{\phi}_{\alpha} + <D, {\phi}_{\alpha}> =
\delta({\varphi}) {[ J_{\nu}^{g} ]}_{\alpha}\\
\label{rooteq}
\eeq

where $J_{\nu}^{\tilde g}$ is simply $Ad_{{\tilde
g}(0)}^{*}(J_{\nu})$,
$[J]_{\gamma}$ denotes the Cartan's part of $J$ and

$[J]_{\alpha} = <J,e_{\alpha}>$.

 From the equation (\ref{careq}) we deduce that $D = constant$ and
 $[J_{\nu}^{g}]_{\gamma} = 0$.
This implies, that be Cartan-valued constant conjugation we can twist
$J^{g}_{\nu}$ to the $J_{\nu}$ itself.
 Then, (\ref{rooteq}) implies that locally (at $\varphi \neq 0$ )
we can represent ${\phi}_{\alpha}({\varphi})$ as follows:

\beq
{\phi}_{\alpha}({\varphi}) =
\exp ( - \frac{\varphi}{\kappa} <D, {\alpha}> ) \times M_{\alpha}
\label{rootres}
\eeq

 where $M_{\alpha}$ is locally constant vector in $g$. Look at the
right hand
side of (\ref{rooteq}) leads us to the conclusion, that $M_{\alpha}$
jumps
when ${\varphi}$ goes through $0$. This jump is equal to

\beq
[\exp ( - \frac{2\pi}{\kappa} <D, {\alpha}> ) - 1] \times M_{\alpha}
= [J^{g}_{\nu}]_{\alpha}
\label{jump}
\eeq

The final answer for the reduction is the following:
the physical degrees of freedom are in the
$\exp(-\frac{2\pi i}{\kappa}X)$
and $P$ (note that $P$'s entries are purely imaginary) with the
reduced
symplectic structure:

\beq
\omega = \frac{1}{2 \pi i} {\sf tr} ( \delta P \wedge \delta X )
\label{redsym}
\eeq

and we have

$$
{\phi}_{\alpha}({\varphi}) = {\nu}
\frac{\exp ( - \frac{i\varphi}{\kappa} <X, {\alpha}> )}
{\exp ( - \frac{2\pi i}{\kappa} <X, {\alpha}> ) - 1}
$$

Now if we would take some simple hamiltonian system on the
$T^{*}{\hat g}$
whose
Hamiltonian is invariant under (\ref{adj}),(\ref{coadj}) actions,
then we will
get somewhat complicated system on the reduced symplectic manifold,
i.e. on
$T^{*}{\bf T}$.
 Collecting together the term which provides the Poisson bracket
,Gauss law constraint with the Lagrangian multiplier $A_{0}$ and
quadratic Casimir

\beq
{\cal H}_{2} = \frac{1}{4\pi} \int_{S^{1}} d{\varphi} <{\phi},
{\phi}>
\label{2cas}
\eeq

we find ourselves with the YM action with additional Wilson line.
On the reduced manifold we are left with the Hamiltonian

\beq
H_{2} = -\frac{1}{2} {\sf tr} P^{2} +
\sum_{\alpha \in \Delta_{+}} \frac{{\nu}^{2}}
{sin^{2}<X,{\alpha}>}
\label{redham}
\eeq

which coincides with the Hamiltonian of the Sutherland model.
Here $\Delta_{+}$ denotes the set of positive roots
of $g$. For example, for $G = SU(N)$ we have the Hamiltonian
for pair-wise interaction between
the particles with the potential:
$$
V_{ij}^{A} = \frac{g^{2}}{sin^{2}(x_{i}-x_{j})}
$$

for $G = SO(2N)$ we have:
$$
V_{ij}^{D} = g_{2}^{2} [ \frac{1}{sin^{2}(x_{i}-x_{j})} +
\frac{1}{sin^{2}(x_{i}+x_{j})}] +
$$
$$
+ g_{1}^{2} [\frac{1}{sin^{2}(x_{i})} + \frac{1}{sin^{2}(x_{j})}]
$$

where $g_{1}, g_{2}$ are coupling constants. In general
case one has as many different coupling constants as many
orbits in
the root system has a Weyl group, i.e. two or one \cite{olsh-per-cl}.

In our situation we have to attach a representation $R_{\nu}$ to the
outgoing end of the Wilson line and the dual to the incoming one.
So, the boundary conditions we
should fix are: vectors $<v_{1}| \in R_{\nu}^{*}, |v_{2}> \in
R_{\nu}$ and the
monodromies
$ g_{2} = P\exp \int_{0_{2}} A$, $ g_{1} = P\exp \int_{0_{1}} A $
around
the initial and final holes respectively.

Gauge transformations will rotate
monodromies and vectors as follows:
$g_{i} \rw h g_{i} h^{-1}, |v_{i}> \rw T_{R_{\nu}}(h) |v_{i}> \;
i = 1,2$ and the answer will be invariant with
respect to these transformations.
The measure on $g_{i}$ is the Haar measure. Using this
gauge freedom, we can make $g$ to be in the Cartan subgroup ${\bf T}
\in G$.
The measure
on ${\bf T}$ induced from the modding out angle variables is
the product of the
Haar measure on the ${\bf T}$ - $\prod_{i} d\theta_{i}$ and
corresponding
group-like VanderMonde determinant.
To get the proper answer for the transition amplitude we should
take a square root of this measure (to get a half-form).
 Now let us turn to the calculation. Let us cut our cylinder along
 the contour $\Gamma$:
we will get the disk, and on the edges, corresponding to $\Gamma$ we
have
some group element $h \in G$, which should be integrated out with the
weight
$<v_{1}|T_{R_{\nu}}(h)|v_{2}>$.
 The whole monodromy around the disk is given by
$g_{1}hg_{2}^{-1}h^{-1}$,
so the answer for the transition amplitude (or evolution kernel) is
the following:
\beq
<g_{2};v_{2}|\exp[-tH]|g_{1};v_{1}> =
\label{kernel}
\eeq
$$
= \sum_{\hh} d_{\hh} e^{-tQ_{2}(\hh + \rho)} \int dh
{\chi}_{\alpha_{\hh}}(g_{1}hg_{2}^{-1}h^{-1})
<v_{1}|T_{R_{\nu}}(h)|v_{2}>
$$

{}From this expression we can extract the spectrum of the model as
well as the
structure of the Hilbert space

$$
{\cal H} = \bigoplus_{\hh} Inv (\alpha_{\hh} \otimes
{\alpha_{\hh}}^{*}
\otimes R_{\nu}) =
$$
$$
= \bigoplus_{\alpha} \sum_{\Phi_{\hh}} {\bf C} \cdot \Phi_{\hh}
$$
where $\Phi_{\hh}: \alpha_{\hh} \to \alpha_{\hh} \otimes R_{\nu}$
is an intertwiner.

$$
E_{\alpha_{\hh}} = \frac{1}{2}<\hh + \rho, \hh + \rho>
$$

{}From the answer for the kernel we can realize that more restrictive
conditions on coupling $\nu$ should
be imposed to be possible to get the Sutherland system
as quantum reduction:
if we take $g_{1},g_{2}$ to be ${\bf T}$ - valued, then the integral
over $h$ will invariant
under left and right independent multiplication of $h$ by
the elements of ${\bf T}$ -
this lead to the condition $T_{R_{\nu}}({\bf T})|v_{i}> = |v_{i}>$.
It is the
integral $\nu$, for which $R_{\nu}$ contains such a vector Moreover,
there exist only one such a vector. Let us denote it as $|0>$.

To get the expression for the wavefunction, let us look once again on
the
integral over $h$.
It can be taken by representing character in the orthonormal base of
representation $\alpha_{\hh}$,
so we will arrive to the expression:

\beq
\Psi_{\alpha_{\hh}}( \{ \theta_{i} \} ) = {\cal N} (\{ \theta_{i} \})
\sum_{mn} C_{mn} <m|T_{\alpha_{\hh}}(diag[e^{{\sf i}\theta_{i}}])|n>
\label{wf}
\eeq

here ${\cal N} (\{ \theta_{i} \})$ denotes the normalization factor,
which comes
from
the taking into account the group-like Vandermonde determinant,$m,n$
run over
the orthogonal base
of the representation $\alpha$  and coefficients $C_{mn}$ are defined
via:

$$
\int dh T_{\alpha}(h)_{mk} T_{\alpha}(h)_{nl}^{*} T_{R_{\nu}}(h)_{00}
=
C_{mn} C_{kl}^{*}
$$

It is instructive to derive an expression for the wavefunctions of
Calogero model via the
large ${\kappa}$ limit in those for Sutherland model. We start from
the
expression
for the transition kernel

$$
{\cal K} (g_{2}, v_{2} | g_{1}, v_{1} ) = \sum_{\alpha} \exp( -
\frac{T}{{\kappa}^{2}} Q (\alpha )) {\cal K}_{\alpha} (g_{2}, v_{2} |
g_{1},
v_{1} )
$$

\beq
{\cal K}_{\alpha} (g_{2}, v_{2} | g_{1}, v_{1} ) =
 \int_{SU(N)} dh \chi_{\alpha} ( h g_{1} h^{-1} g_{2}^{-1}) <v_{2}|
T_{R_{\nu}}(h) |v_{1}>
\label{amplit}
\eeq

Let us write $g_{1,2} = \exp ( \frac{i}{\kappa} Q_{1,2})$, then
$ h g_{1} h^{-1} g_{2}^{-1} = \exp ( \frac{i}{\kappa} \Phi )$ with

$$
\Phi = h Q_{1} h^{-1} - Q_{2} + {\cal O}(\frac{1}{\kappa})
$$

Now we can write down an expression for the eigenvalues of the matrix
$h g_{1}
h^{-1} g_{2}^{-1}$ as $e^{\frac{i\Phi_{j}}{\kappa}}$ where $\Phi_{j}$
are
eigenvalues of $\Phi$. By the Weyl character formula we have:

\begin{eqnarray}
&& dim ( \alpha) = \frac {\Delta (\ppb)}{\Delta (i)}
\nonumber\\
&& \chi_{\alpha}(U) = \frac{det||{\beta}_{j}^{\ppb}||}{{\Delta}
(\beta_{j})}
\label{irreps}
\end{eqnarray}
where $U $ is conjugate to $diag(\beta_{1},\ldots,\beta_{N})$.
Here  collection of $\ppb$'s is the highest weight $\hh$ of the
irreps $\alpha_{\hh}$.

The formula we
will derive uses the representation of the large $\kappa$ limit
of
the normalized character through
the Harish-Chandra(Itsykson-Zuber) integral,
namely, if  $\frac{\hh}{\kappa} = {\bar P}$   is fixed, then

$$
\lim_{{\kappa} \rw { \infty}}\frac{\chi_{\alpha}(U)}{dim(\alpha)} =
\int_{SU(N)} dh \exp ( i Tr ({\bar P} h{\bar Q}h^{-1})
$$
where $\exp (i\frac{\bar Q}{\kappa})  = diag ( U )$, and $diag (U) $
denotes
any diagonalized form of the matrix $U$ . Substituting this
expression into
the integral in (\ref{amplit}), we get an integral of the type:

\begin{eqnarray}
&& {\cal K}_{\bar P} (g_{2}, v_{2} | g_{1}, v_{1} ) =
\nonumber\\
&& \Delta ({\bar P}) {\kappa}^{N(N-1)/2} \times
\nonumber\\
&& \int\int_{SU(N)}dh dh^{\prime} \exp ( i Tr (h^{\prime} (h Q_{1}
h^{-1} -
Q_{2}) {h^{\prime}}^{-1} {\bar P})
\nonumber\\
&& <v_{2}| T_{R_{\nu}}(h) |v_{1}>
\label{limamplit}
\end{eqnarray}
Finally, we can extract a wavefunction from the condition:

$$
{\Psi}_{\bar P}^{\dag} ({\bar Q}_{2}) {\Psi}_{\bar P} ({\bar Q}_{1})
=
\int\int_{SU(N)} dh dh^{\prime} {\cal K}_{\bar P} (g_{2}^{h},
v_{2}^{h} |
g_{1}^{h^{\prime}}, v_{1}^{h^{\prime}} )
$$
where $g^{h} = h g h^{-1}$, $v^{h} = T_{R_{\nu}}(h)v$.

It yields essentually :

$$
{\Psi}_{\bar P} ({\bar Q}) ={\Delta}({\bar P}) {\Delta}({\bar Q})
\int_{SU(N)} dh \exp ( i tr
({\bar P} h {\bar Q} h^{-1}) <0| T_{R_{\nu}} (h) |0>
$$

   In the case of $N=2$
this integral reduces to the integral representation for the Bessel
function.  It is interesting to note, that the initial expression of
the
wavefunction for the Sutherland model could be regarded as the same
formula
applied to the central extended loop group -- in that case we would
get some
path integral, which we can evaluate using, for example, localization
technique, or, by choosing appropriate coordinates on the co-adjoint
orbit
of affine algebra.

\section{Deformation of CSM models - trigonometric Ruijsenaars's
system.}
\setcounter{equation}{0}

In this section we will make the next step and consider the
deformation of the previous
constructions.The subject of deformation is the phase space.According
to the procedure described above  the each phase space gives rise to
the particular topological theory like topological YM theory in the
previous section.Now the theory at hands is G/G WZW theory and the
full action to be considered is the generating functional for G/G
theory with additional Wilson line.

Therefore, let us consider the Hamiltonian reduction of the cotangent
bundle to
the central extended loop group $\hat G$. This symplectic manifold is
a
space of 4-tuples

$(g:S^{1} \rw G, c \in U(1) ; A \in \Omega^{1}(S^{1}) \otimes g^{*},
{\kappa}
\in {\bf R})$.

The action of an element $h \in {\cal L}G$ is the following:

$$
g \rw h g h^{-1}, \; A \rw h A h^{-1} + {\kappa} h {\partial} h^{-1}
$$

$$
{\kappa} \rw {\kappa}, \;
c \rw c \times {\cal S}( g, h)
$$

where $\cal S$ is constructed from the $U(1)$-valued two-cocycle on
the group
${\cal L}G$,
$\Gamma (g, h)$
$$
{\cal S} (g, h) = \Gamma (h, g) \Gamma(hg, h^{-1})
$$

As always, on this cotangent bundle exists a natural symplectic
structure and
the
loop group action preserves it.
 The symplectic form $\Omega$ has the
following structure:
\begin{eqnarray}
 && \Omega = \int_{S^{1}} {\sf tr}
[
A (g^{-1} \delta g)^{2}
+ \delta A \wedge
g^{-1} \delta g ] +
\nonumber\\
&& \int_{S^{1}} {\sf tr} [ {\kappa}
{\pdv}g \cdot g^{-1} (\delta g \cdot g^{-1})^{2} -
\kappa \delta ({\pdv}g) \cdot g^{-1} \delta g \cdot g^{-1}] +
\nonumber\\
&& + c^{-1} \delta c \wedge \delta \kappa
\label{loopsympl}
\end{eqnarray}

According to the general scheme we introduce the moment map which
serves as the generalization of the Gauss law in YM theory.
The moment map of the action of loop group
has the form:

\beq
\mu (g,c; A,\kappa) = (g A g^{-1} + {\kappa} g {\partial} g^{-1} - A,
0)
\label{momgru}
\eeq

It is tempting to make the Hamiltonian reduction under some
appropriate
level of
the moment map. It is clear that in this way we deform Sutherland
system,
 preserving its integrability. For simplicity we consider only
$SU(N)$ case.
The reasoning, similar to that in the YM case, suggests
the level of moment map to be

\beq
\mu (g,c; A,{\kappa}) = {\sf i} \nu (\frac{1}{N}
Id - e \otimes e^{+}) \delta (\varphi)
\label{momlev}
\eeq

As we did it in the previous sections, by general gauge
transformation
$H(\varphi)$
we can make $A$ to be constant diagonal matrix $D$, defined modulo
affine
Weyl group action, i.e. modulo permutations and addings of the
integral
diagonal matrices. Generally, such a transformation
doesn't respect the value of the moment map. Let $H$ denotes the
value of
$H(0)$
at the point $0$. The matrix $H$ defines a point in the coadjoint
orbit of
$J$,
i.e. ${\cal O}_{J} ={\bf CP}^{N-1}$. The semi-direct product
of Cartan torus $U(1)^{N-1}$ and Weyl group ${\cal S}_{N}$ acts on
${\cal O}_{J}$ by permutations of homogeneous coordinates and
their multiplications by phases.

The equation (\ref{momgru}) takes the form:

\beq
g D g^{-1} + {\kappa} g d g^{-1} - D =
{\sf i} \nu (\frac{1}{N} Id - f \otimes f^{+})\delta(\varphi)
\label{mastereq1}
\eeq

Here $f = H e$ is some vector in ${\bf C}^{N}$ with unit norm $<f,f>
= 1$.
We can assume that $f \in {\bf R}^{N}$, due to the remaining
possibility of
the left multiplication $H \rw Y H$ with $Y$ being diagonal unitary
matrix
(it comes from the gauge freedom which survives after diagonalization
of
$A$). Equation (\ref{mastereq1}) yields immediately:

$$
g = \exp ( \frac{\varphi}{\kappa} D) G(\varphi)
\exp( -\frac{\varphi}{\kappa} D) ;
\partial_{\varphi} G = - \frac{J}{\kappa} G \delta(\varphi) $$

where $J = {\sf i} \nu ( \frac{1}{N} Id - f \otimes f^{+} )$.
Let us also introduce a notation for the monodromy
of connection $D$: $Z = \exp ( - \frac{2\pi}{\kappa} D) = diag(z_{1},
\dots,
z_{N})$,  $\prod_{i} z_{i} = 1$,
$z_{i} = \exp (\frac{2{\pi}i q_{i}}{\kappa})$.
We have a boundary condition:

\beq
 {\tilde G}^{-1} Z {\tilde G} = \exp(\frac{2\pi J}{\kappa}) Z
\label{commutant}
\eeq

where $\lll = e^{\frac{2 \pi i\nu}{N \kappa}}$,$ {\tilde G} = G(+0)$.

 It turns out that solution has the following form: let
$P(z)$ will denote a characteristic polynomial of $Z$, i.e.  $$ P(z)
=
\prod_{i} (z - z_{i}) $$

Let
$$
Q^{\pm}(z) =
\frac{P({\lll}^{\pm 1}z) - P(z) }{({\lll}^{\pm N}-1)z P^{\prime}(z)}
$$
- the ratio of the finite difference and derivative of $P$.
When $\lll \to 1$,
the rational functions $Q^{\pm}(z)$ tend to $\frac{1}{N}$.

In these notations the matrix $\tilde G$ can be written as follows:

\beq
{\tilde G}_{ij} = - \lll^{-\frac{N-1}{2}}
\frac{{\lll}^{-N}-1}{{\lll}^{-1}z_{i}-z_{j}}
e^{i\theta_{i}}(Q^{+}(z_{i})Q^{-}(z_{j}))^{1/2}
\label{Lax}
\eeq

$$
=e^{i\h_{i}-\frac{\pi i}{\kappa} (q_{i}+q_{j}) }
\frac{sin(\frac{\pi\nu}{\kappa})}{sin(\frac{\pi(q_{ij}-
\frac{\nu}{N}}{\kappa})}
\prod_{k \neq i, l \neq j}
\frac{sin(\frac{\pi{q_{ik} +
\frac{\nu}{N}}}{\kappa})}{sin(\frac{{\pi}q_{ik}}{\kappa})}
\frac{sin(\frac{{\pi}{q_{il} -\frac{\nu}{N}}}{\kappa})}{sin(
\frac{{\pi}q_{il}}{\kappa})}
$$

where
$\h_{i}$ are some
undetermined phases,
which are going to be the momenta,
corresponding to the coordinates  $q_{i}$.
It means, that on the reduced symplectic manifold the
symplectic structure is $\sim \sum_{i} d\h_{i} \wedge dq_{i}$.

The next step is to consider some simple Hamiltonian system on
$T^{*}{\hat G}$, which is invariant with respect to the loop group
adjoint
action and make a reduction. It is obvious that any $Ad$-invariant
function
$\chi : G \rw {\bf R}$ defines a Hamiltonian

$$
H_{\chi} = \intt d{\varphi} {\chi}(g)
$$

For example, $\chi_{\pm} (g) = {\sf Tr} (g {\pm} g^{-1})$ (here $\sf
Tr$
is a
trace
in the $N$-dimensional
fundamental
representation of $SU(N)$ )
give the Hamiltonians (up to a constant factor, independent of
$q_{i}$) :

\beq
H_{\pm} = \sum_{i} (e^{i \h_{i}} {\pm} e^{- i\h_{i}})
\prod_{j \neq i} f(q_{ij})
\label{ruuham}
\eeq

where the function $f(q)$ is given by:
$$
f^{2}(q) = [ 1 -
\frac{sin^{2}
({\pi}{\nu}/{\kappa} N)}
{sin^{2}({\pi}q/{\kappa})}
].
$$
One recognizes her the Hamiltonian of the Ruijsenaars system.
For detailed elaboration of different limits of the Hamiltonian
see, for example \cite{ruu}.  In particular, taking an appropriate
scaling
limit $\kappa \to \infty$ will return us back to Sutherland model.
Note also, that in order to have a real-valued Hamiltonian we need
$\frac{\nu}{N} < q_{ij}$
for any $i{\neq}j$,
thus $\frac{\nu}{\kappa}$ should be
restricted from above.This restriction is wellknown for the
integrable representations of KM algebra.It can be expected that some
kind of new phenomena happens above this value of the coupling
constant in the system of the particles.

Now we are ready to answer the question what field theory Ruijsenaars
system corresponds to.
Action will be a sum of the term like $\int pdq$ and a moment value
fixing
term $\int A_{t} \mu dt$, where $A_{t}$ is going to be a time-like
component of a gauge field, serving as a Lagrange multiplier.
Thus, the action is (we omit the term $c^{-1}{\partial_{t}}c \kappa$,
while fix the non-vanishing level $\kappa$):

$$
S( A, g)
= \int d{\varphi}dt
{\;\sf tr} [
- A_{\varphi} g^{-1} {\partial_{t}} g
- \kappa {\partial_{t}}g \cdot g^{-1} \cdot {\pdv} g \cdot
g^{-1}
 +
$$

$$
+ \kappa d^{-1}
({\pdv} g \cdot g^{-1} (dg \cdot
g^{-1})^{2}) +
$$

$$
+ A_{t} ( \kappa g^{-1}
{\pdv} g + g^{-1} A_{\varphi} g -
A_{\varphi} )]
$$

This is the action of $G/G$ gauged WZW model \cite{gwzw}, a
topological
gauge theory recently shown to give a Verlinde formula for the
dimension of
the space of conformal blocks in WZW theory on the surface \cite{bt +
ger}.

Our theory will contain also a Wilson line in the representation
$R_{\nu}$.and the Hamiltonian written above. To compute the path
integral on a cylinder, as in the case
of Yang-Mills theory,one cuts a cylinder along the Wilson line and
the
answer for the evolution cernel will have the form (\ref{kernel}).In
fact
the representations will run only through integrable representations
of
$SU(N)_{k}$. We conjecture that the answer can be obtained also from
$U_{q}(SL_{N})$ representation theory, thus establishing the
relation between this quantum group and gauged WZW theory once
again.Let us emphasize that the reduced phase space coincides with
the moduli space of the flat connections on a punctured torus so once
again we have the dynamics on the moduli space.

It is useful to present the interpretation of the previous
construction
in terms of
Chern-Simons field theory,
deformed by some Hamiltonian. Let us consider a
Chern-Simons theory with gauge group $SU(N)$ on the space-time
manifold
 being a product
of the interval and a two-torus $X = I \times T^{2}$. Its action is
given by:

\beq
S_{CS} = \frac{i {\kappa}}{ 4\pi}\int_{X} {\sf tr} (A \wedge dA +
\frac{2}{3} A \wedge A \wedge A)
\label{CS}
\eeq

We know that the phase space of Chern-Simons theory on a
${\Sigma} \times I$ with
$I$ being an interval is a moduli space of flat connections. In the
case of
inserted into the path integral Wilson lines, let us say just one
Wilson
line in the representation $R$,  when the integral in hands looks
like:

$$
{\ddd} A <v_{1}| T_{R}(P \exp \int A ) |v_{2}> \exp( - S_{CS}(A) )
$$
the answer will be slightly changed, namely we will have to consider
a
moduli space of connections on the punctured surface
with prescribed conjugacy class of monodromy of connection
around the puncture. This class $U$ is
 related to the highest weight $\hat h$ of the representation
$R_{\nu}$ like
$ U = \exp ( \frac{2\pi {\sf i}}{{\kappa} + N} \mbox{diag} ( {\hat
h}_{i}))$
If our surface is a two-torus $\Sigma = T^{2}$, then we can write
immediately the condition on the monodromies $g_{A}, g_{B}$ of
connection
around $A$ and $B$ cycles and the monordomy $g_{C}$ around the
puncture :
$$
g_{A} g_{B} g_{A}^{-1} g_{B}^{-1} = g_{C}
$$
Keeping in mind
the condition on the conjugacy class of $g_{C}$, taking into
account that for the representation $R_{\nu}$ we have a signature
like ${\nu}
\times \mbox{diag}
(N, 0, \dots, 0)$ we arrive to the equation
(\ref{commutant}).
Also we could point out the
equivalence of
the sector of the space of
observables in the
Chern-Simons theory on
${\Sigma} \times S^{1}$ ,
namely, observables,
which depend on
$A_{t}$ with ${\partial}_{t}$
being a direction, tangent to $S^{1}$ and the gauged $G/G$ WZW
theory.
This can be proved either considering a gauge fixing
$A_{t} = \mbox{const}_{t}$ and diagonal,
or by poining out that
the wave functions in
the Chern-Simons theory are
those in the theory with an
action
$\int dt \int_{\Sigma} {\sf tr} (A {\partial_{t}} {\bar A})$
projected onto the space of functionals,
invariant under the gauge group action.
This involves taking into
account the non-invariance of
the polarization
$\frac{\delta}{\delta{\bar A}}$
and this is why the
WZW part of
the action appears (see \cite{bt + ger} for detailed discussion).

{}From this equivalence we can derive a spectrum of the
Ruijsenaars model.
Namely we will show that it is contained in the spectrum of
the free relativistic fermionic system on the circle.
By free relativistic
model of fermions on
the circle we
mean the system of
particles on
the circle with coordinates
$0 < q_{i} \leq 1$ and conjugate momenta
(rapidities) ${\h}_{i}$ living also on the circle, with the
Hamiltonian
$$
H_{+} = \sum_{i} cos (\h_{i})
$$
and symplectic structure
$$
\omega =\frac{1}{2\pi} ({\kappa} + N) \sum \delta \h_{i} \wedge
\delta q_{i}
$$
(in fact, the shift ${\kappa} \to {\kappa} + N$
is a quantum effect
and can be deduced from
\cite{bt + ger}, so we don't consider its derivation here).
The
fermionic nature of these particles
amounts to the phase space
$({\bf T}^{N-1} \times {\bf T}^{N-1})/ {\cal S}_{N}$ -
just the moduli space of flat connections on
the torus. Here ${\bf T}^{N-1}$ is a Cartan torus for
$SU(N)$ (center of mass is fixed) and ${\cal S}_{N}$
is a symmetric group.
We call them fermions,
because  their wavefunction vanishes on the diagonals $z_{i} = z_{j}$
in the $N-1$-torus (it contains a factor like
$\Delta ( e^{{\sf i}\h_{i}} )$ - a group-like Vandermonde).
It is obvious,
that
the corresponding Schr\"odinger equation will be a difference
equation, because
$cos( \frac{2\pi{\sf i}}{{\kappa} + N} {\partial}_{q})$
is a finite difference operator.
Its eigenfunctions on the circle are
exponents $e^{2 {\pi}{\sf i}n q}$ ($n$ is necessarily integral) with
eigenvalue $$ E_{n} = cos (\frac{2\pi n}{{\kappa} +N}) $$ and the
total
spectrum will be given by the sum $$ \sum_{i} E_{n_{i}} $$ where we
have to
impose the following conditions, coming from the invariance $E_{n} =
E_{n +
{\kappa} + N}$ and the symmetric group action:  \beq ({\kappa} + N )
>
n_{N}> \dots > n_{i} > \dots > n_{1} \geq 0 \label{restr1} \eeq \beq
\sum_{i} n_{i} < ({\kappa} + N)
\label{restr2}
\eeq

In our situation fermions on the circle are not
completely "free" and the spectrum
is restricted. For our purposes will be more
convenient to form a vector
$\hh = (n_{i} - i), i=1,2,\dots,N$. It will correspond to the
dominant weights of the $SU(N)$ irreducible representations.
Now we can write an answer for the  spectrum:
$$
E_{\hh} = \sum_{i} cos(\frac{2\pi \hh_{i}}{\kappa + N})
$$

and $\hh$ satisfy the selection rule: there exists non-trivial
intertwiner

$$
\Phi_{\hh} : \alpha_{\hh} \to \alpha_{\hh} \otimes R_{\nu}
$$

(essentially it implies, that $\hh = \lambda + \nu \rho$, where
$\lambda_{i} \geq 0, i=1, \dots, N$.

An explanation of
this spectrum can be
derived from the properties of gauged $G/G$ WZW model \cite{bt +
ger}.
Namely,
to get a
partition
function of
our interest
we have to compute a path integral on the two-torus:
\beq
\int
{\ddd}A{\ddd}g
\exp(
S_{GWZW}(g, A) +
\int_{T^{2}} {\sf Tr} ( g + g^{-1}) )\chi_{R_{\nu}} ( P \exp \int_{C}
A )
\label{masterpath}
\eeq
where
$C$ is a
non-contractible
contour on the torus,
$\chi_{R_{\nu}}$
is a character in
the representation $R_{\nu}$.
By the standard arguments we can calculate
it by
cutting the torus along
$C$ and computing first
the path integral in the theory
$S_{GWZW}(g, A) + \int {\sf Tr} ( g + g^{-1}) $
on the annulus with coinciding monodromies $h$ of
the restrictions of
the
connection
$A$ on
the boundaries
of the annulus and
then integrate the answer over $H$ with the weight
$\chi_{R_{\nu}}(h)$.
  In arbitrary gauge theory in two dimensions the
path integral on the disk and
on the cylinder can be written in terms of sum over all
irreducible representations of the gauge group via the characters
in these representations.

Taking into
account Wilson line in the representation $R_{\nu}$ will lead
us to Macdonald polynomials  which play the role of zonal spherical
functions on the quantum group.

 In fact, we have two important parameters in hands:
$$
q = e^{\frac{2\pi i}{\kappa + N}}
$$
and
$$
t = q^{\nu + 1}
$$

. These polynomials
have very interesting limits (for example,
they are responsible for the zonal spherical
functions on the p -adic groups,namely Mautner-Cartier polynomials
when
$q=0,t=1/p$) if $q,t$ tend to some exceptional values.
Unfortunately, in
our approach $q$ should be a root of unity and $t$ should be an
integral
power of $q$, so there is no obvious way how to recognize all this
beauty of
Macdonald polynomials in gauged $G/G$ WZW theory for $SU(N)$
group.Note that path integral representtation from this section
allows to derive the integral representation for Macdonald
polynomials \cite {nikita}.

It is known that Ruijsenaars system is closely related with the
solitonic
sector of Toda like theories,in particular asymptotics of the wave
functions
provides S-matrix for scattering in the solitonic sector
\cite{solitons}.
The qualitative explanation of this relation looks as follows.
Lagrangian
approach for Toda theories is based on GWZW actions with some
additional
dependence on the spectral parameter. In the consideration above we
added
Hamiltonians and took some peculiar  level of the momentum map. Thus
the final action is nothing but the generating functional for GWZW
while
the level of the momentum map fixes the particular solitonic
solutions.
Hence we considered in essence the generating functional and
solitonic
S-matrix naturally appears in this context. Certainly this relation
needs
for further clarification, especially in context of (q)KZ equations
describing the solitonic scattering amplitudes

\section{Elliptic Calogero model}
\setcounter{equation}{0}

Another generalization of  the trigonometric model leads us to the
elliptic system which has a new features.At first it is an example of
the integrable system
 with the spectral parameter which lives on the elliptic
curve.Secondly it naturally arises from 3D field theory which will be
presented below.The phase space we start with now is the cotangent
bundle to the Lie algebra of

$SL(N,C)$ valued functions on the elliptic curve with modular
parameter
$\tau$.

Let us take a consider
cotangent bundle $T^{*} {\elg}$ , which is a space of 4-tuples
$$
( \phi, c ; {\kappa}{\bar \partial} + {\bar A}),

\; \phi :  \curve \to \slnc, \; {\bar A} \in
\Omega^{(0,1)}(\curve) \otimes \slnc, \; c, \kappa \in {\bf C}
$$
the pairing between algebra $\elg$ and its dual $\eld$ in this
parameterization has a form
\beq < (\kappa {\bar \partial} + {\bar A}); \; (\phi, c) > =
\kappa \cdot c + \intc \omega \wedge tr \phi {\bar A}
\label{pairing}
\eeq
where $\omega$ is a holomorphic 1-differential with periods along A
and B cycles 1 and $\tau$ correspondingly.

On the cotangent bundle acts naturally a current group ${\it
SL}_{N}({\bf
C})^{\curve}$:
\begin{eqnarray}
&& \ppz \to \ggz \ppz \ggz^{-1}
\nonumber\\
&& \aaz d{\bar z} \to  \ggz \aaz \ggz^{-1} + {\kappa}
\ggz {\bar \partial} \ggz^{-1}
\nonumber\\
&& \kappa \to \kappa, \; c \to c + \intc \omega \wedge

tr (\phi g {\bar \partial} g
)
\label{graction}
\end{eqnarray}

This action preserves a holomorphic

symplectic form $\Omega$ on $T^{*}{\elg}$:

\beq
\Omega = \delta c \wedge \delta \kappa +
\intc \omega \; tr ( \delta \phi \wedge \delta {\bar A} )
\label{sympl}
\eeq

and defines a  moment map:
\beq
\mu = {\kappa} {\bar \partial}\phi + [{\bar A} , \phi ]
\label{moment}
\eeq

It will be convenient to enlarge this symplectic manifold by the

finite-dimensional $\SLnc$ - coadjoint orbit ${\cal O}_{\nu}^{-}$

. The sign $-$ denotes the reversed

symplectic form on ${\cal O}_{\nu}$. The enlarged
symplectic manifold
$$
X_{\nu} = T^{*}{\elg} \times {\cal O}_{\nu}^{-}
$$

The moment map $M$ for the current group action on $X_{\nu}$ is

$$
M = {\kappa}{\bar \partial} {\phi} + [{\bar A}, {\phi} ] - J {\delta(
z, {\bar z}) \frac{dz \wedge d{\bar z}}{\omega},
$$
for the point in the coadjoint orbit $J \in {\cal O}_{\nu}$

Now we wish to apply a hamiltonian reduction at zero level of moment
map.
It means that we have to impose a constraint

$$
M = 0
$$
and  take a quotient of the submanifold $M^{-1}(0)$ by the action of
the

current group.

It amounts to
$$
0 = {\kappa}{\bar \partial} {\phi} + [{\bar A}, {\phi} ] - {\nu}({\bf
1} -

u \otimes v^{*})\frac{\delta (z, {\bar
z}) dz \wedge d{\bar z}}{\omega}
$$

The generic element of $\elg$ by the action (\ref{graction}) can be
transformed
to semi-simple constant element of maximal torus of $\slnc$. After
applying this gauge transformation we arrive to equation, written
in terms of matrix elements:
\beq
{\kappa} {\bar \partial}\phi_{ij} + a_{ij} \phi_{ij} = {\nu}
(\delta_{ij} - u_{i}v_{j}^{*} ) {\delta}(z,{\bar z}),
\label{***}
\eeq
here $a_{ij} = a_{i} - a_{j}$, $a_{i} \in
{\bf C}$ are the diagonal entries of $\bar A$.

 Equation (\ref{***}) yields:

{\bf a.)} $ 1 - u_{i} v_{i}^{*} = 0 $

{\bf b.)} $\phi_{ii} = p_{i} = const, i = 1, \dots, N ; \sum_{i}
p_{i} = 0$

{\bf c.)} $\phi_{ij} = - \frac{2{\pi}{\sf i}{\nu}}{\kappa}
u_{i}/u_{j}
\exp( \frac{a_{ij} (z - {\bar z})}{\kappa} )

\frac{{\haha}(z + \frac{{\pi}a_{ij}}{{\kappa}{\tau_{2}}})}{{\haha}(z)
{\haha}(\frac{{\pi}a_{ij}}{{\kappa}{\tau_{2}}})}$, for $i \neq j$,
and  $\tau_{2} = $Im$\tau$, $\haha$ is one of the four
$\theta$-functions,

namely that one, which
vanishes at $0$.

 To prove {\bf c.} we
perform the following substitution
$$
\phi_{ij}(z,{\bar z}) = \exp ( \frac{a_{ij} (z - {\bar z})}{\kappa} )
\psi_{ij} (z, {\bar z})
$$
where $\psi_{ij}$ has non-trivial monodromy properties
$$
\psi_{ij} (z + 1, {\bar z} + 1) = \psi_{ij}(z, {\bar z})
$$
$$
\psi_{ij} (z + \tau, {\bar z} + {\bar \tau}) =

e^{-\frac{2\tau_{2} {\sf i}}{\kappa} a_{ij}}

\psi_{ij} (z, {\bar z})
$$
Equation (\ref{***}) tells us that ${\bar \partial} \psi_{ij} =
-{\nu}

u_{i}/u_{j} {\delta}(z, {\bar z})$, thus $\psi_{ij}$ is a meromorphic
section of the holomorphic line bundle with the prescribed transition
functions
and with the only  residue at $0$ equal $-2\pi {\sf i} \nu
u_{i}/u_{j}$.
The solution is unique and it gives {\bf c.}.

After reduction we are left with the following variables:
$\{ p_{i} \}, \{ a_{i} \}, \{ u_{i} \}$. Moreover,

$\sum p_{i} = \sum a_{i} = 0$. But this is not the end of the story.
In fact, fixing the representative of ${\bar A}$ leaves us with the
residual gauge transformations, generated by constant diagonal
$\SLnc$ matrices. They act non-trivially only on $u_{i}$ and this
action permits us to fix them completely: $u_{i} \to 1$, for any $i$.

In fact, the matrix $\psi$, when evaluated at some point $z \neq 0$
gives

precisely Krichever's answer \cite{krichever}
for the Lax matrix.

Invariants of the matrix $\ppz$ give us the Hamiltonians for
integrable model on $\cal M$. The first non-trivial one is
$$
\frac{1}{2}\mbox{tr} \ppz^{2} = \sum_{i} \frac{1}{2} p_{i}^{2} +
\frac{4{\pi}^{2}{\nu}^{2}}{\kappa^{2}} [\sum_{i < j} {\wp} (
\frac{{\pi}a_{ij}}{{\tau_{2}} \kappa}) -
\frac{N(N-1)}{2} {\wp}(z)]
$$

Thus, we have got a Hamitonian of

elliptic Calogero-Moser model and as always due to
 the quantum corrections coupling constant ${\nu}^{2}$ gets shifted
to
 $\nu(\nu -1)$.

The next non-trivial integral is

$$
\frac{1}{6}\mbox{tr}{\ppz}^{3} = \sum_{i} \frac{p_{i}^{3}}{6}  +
\sum_{i \neq j} \frac{p_{i}}{2} \wp(\frac{
{\pi}a_{ij}}{{\tau_{2}}\kappa}) +
\dots
$$

{}.

As before we expect that some kind of elliptic
deformation of Yang-Mills theory exists.
Here comes the  action of this  theory:
\beq
S_{\tau} = \int_{\curve \times S} \omega \wedge \mbox{tr} (\phi
F_{t {\bar z}} - {\varepsilon} \phi^{2})
\label{ellaction}
\eeq
Here $F_{t {\bar z}} = \partial_{t} {\bar A} - {\bar \partial} A_{t}
+
[A_{t}, {\bar A}]$ is a component of the gauge field strength,
$\omega = dz$ -
holomorphic
differential on $\curve$, $S$ is a time-like circle. It is clear,
that
this theory can be coupled to vertical Wilson lines ($P \exp \int
A_{t}$).
In particular, Wilson line in the representation $R_{\nu}$ produces
our
derivation of elliptic CSM model.
In the limit $\tau_{2} \to \infty$ only those modes of gauge fields
survive
which give rise to the theory on a cylinder,
rather then on a three-dimensional manifold.
Note, that if one adds
Chern-Simons action to the Lagrangian above the system of interacting
particles in the external magnetic field appears.

Note also that elliptic Calogero model is a natural generalization of
the Hitchin,s systems on the moduli space of holomorphic G bundles on
the surface without punctures.

Another popular finite dimensional integrable systems are of the Toda
type.Fortunately elliptic Calogero-Moser system covers both
 periodical and nonperiodical Toda chains and lattices \cite{inoz}
which can
 be considered as a particular limits of the system above. To perform
 reduction to Toda theory one should introduce new variables like
 $a_{i}=x_{i}+(j-1) \frac{b}{\kappa}$ rescale the coupling constant
 $\nu={\nu}_{0}e^{\frac{b}{\kappa}}$ and take the limit $b \to
\infty$.

Let us mention one another important observation. Recall, that the
elliptic Sklyanin algebras \cite{skly2} appeared in description of
asymmetric spin systems with the nearest-neiboughrs interactions.

This
quadratic algebra has a particular representation in which

generators can be
realized as finite-difference operators with coefficients expressed

in terms
of elliptic functions. These operators act in the finite dimensional
representations of $sl_{2}$. The degeneration
procedure of \cite{skly2} yields the differential operators with
elliptic coefficients. The quadratic Casimir in particular
finite-dimensional representation produces the two
particle elliptic Calogero-Moser Hamiltonian.

Let us outline the
difference between interpretation of spectral parameter in spin
systems and
Yang-Mills theory. In spin systems it is the coordinate on the
parameter
space while in YM situation it is nothing but the coordinate on the
world
sheet. The meaning of this difference as well as the role of elliptic
algebra
in deformed YM theory definitely could shed some

additional light on the structure
of the integrable systems in 3d. In particular, it would be
interesting to find in this approach an explanation of the origin
of the generalization of Yang-Baxter equation for elliptic
Calogero-Moser system \cite{sklya}.
It is natural expectation to get elliptic
Ruijsenaars' models \cite{ell ru}, which is the top system for the
tower of
integrable Hamiltonians for interacting particles, as a reduction of
the
integrable system on the cotangent bundle to current group in two
dimensions
(see \cite{gorrui}).

The corresponding topological field theory
will be $3+1$ dimensional and its action can be written as

$$
\int_{\Sigma \times \curve} \omega \wedge [ A dA + {2\over{3}} A^{3}
]
$$
where 'gauge' field $A$ has only three components:
$A_{w}, A_{\bar w}$ and $\bar A$, and

$w,{\bar w}$ are coordinates on the surface

$\Sigma$.

\section{Duality}
\setcounter{equation}{0}
 In this section we show how the transition to the next level in the
hierarchy of the phase spaces related with the finite dimensional
groups can be carried out.The phase space to be treated now is the
Heisenberg double and we show that the corresponding many-body
problem  is the trigonometric Ruijsenaars system once again.Another
subject which will be discussed below is the duality property between
the pairs of the many-body problems.

 To start the consideration of the Heisenberg double which is a pair
$(G,G^{*})$ let us recall, that the group $G^{*}$ is made of the
pairs $(L_{+},L_{-})$  with
the following composition law:
\beq
(L_{+},L_{-})\cdot (M_{+},M_{-}) = (L_{+}M_{+}, L_{-}M_{-})
\label{compos}
\eeq
As a manifolds  $G$  and  $G^{*}$  are  isomorphic  and isomorphism
mapping $G^{*} \to G$ is given by
\beq
(L_{+},L_{-}) \in G^{*} \to L=L_{+}L_{-}^{-1} \in G.
\label{iso}
\eeq
 The following  Poisson brackets  are  multiplicative  on  $G^{*}$
w.r.t. \ref{compos}:
\begin{eqnarray}
&& \{ L_{+} {,}^{\otimes} L_{+} \} =                             -
[r_{\pm}, L_{+} \otimes L_{+}] \\
\nonumber\\
&& \{ L_{-} {,}^{\otimes} L_{-} \} =                             -
[r_{\pm}, L_{-} {\otimes}L_{-}] \\
\nonumber\\
&& \{ L_{+} {,}^{\otimes} L_{-} \} =                             -
[r_{-}, L_{-}{\otimes}L_{+}]\\
\nonumber\\
&& \{ L_{-} {,}^{\otimes} L_{+} \} =                             -
[r_{+}, L_{+}{\otimes}L_{-}]
\label{brackets}
\end{eqnarray}

We use here tensor notations and $r_{\pm} \in g \otimes g$
are classical     $r$-matrices:
$$
r_{\pm}= {\pm} {\frac{1}{2}}
\sum_{i} h_{i} \otimes h_{i} {\pm} \sum_{\alpha \in \Delta_{+}}
e_{\pm \alpha} \otimes e_{\mp \alpha}
$$.

Relations (\ref{brackets}) can be rewritten in a compact form as
the following brackets for $L=L_{+}L_{-}^{-1}$:

\beq
\{ L{,}^{\otimes} L \} =  - (L \otimes 1)r_{+}(1 \otimes L) -
(1 \otimes L)r_{-}(L \otimes 1) +  (L \otimes L)r_{\pm}
+ r_{\mp}(L \otimes L).
\label{lbrackets}
\eeq

  Heisenberg double $D_{+}$ is isomorphic to $G \times G$ and  is
a Lie  group  with a natural component-wise multiplication but it
is not  a  Poisson-Lie  group   w.r.t. the  Poisson   structure   on
$D_{+}$. The Poisson structure is described  in  a
following manner \cite{almal}.
 Let us parametrize an element $L \in G$ with the help
of a Cartan-valued  matrix $C$, $L = g^{-1}Cg$. The symplectic form
can be presented in the form

\beq
\Omega (u,v,C) =\vartheta(u,C) +\vartheta(v,C^{-1})  +  Tr\delta
CC^{-1}{\wedge}(\delta uu^{-1} - \delta vv^{-1}),
\label{sympldouble}
\eeq

where $g   =   v^{-1}u$   and  $ \vartheta(g,C)    =    1/2Tr(\delta
L_{+}L_{+}^{-1} - \delta L_{-}L_{-}^{-1}){\wedge}g^{-1}\delta g$.
In what follows we impose the constraint
\beq
g^{-1}LgL^{-1} = e^{J},
\label{constraint}
\eeq
where the element $e^{J} \in G^{*}$.

The Poisson  brackets  (\ref{lbrackets})  are relevant to
$r$-matrix structure on the space of lattice connections with the
values in
the group $G$ on the torus with one marked  point \cite{fro}. The
latter  can  be
viewed as a bunch of two circles with only one joint point (different
from
the marked one).
Then the  symplectic form (\ref{sympldouble}) is interpreted as a
direct sum of     two     type      summands:
$  \vartheta(u,C)      +
\vartheta(v,C^{-1})$, which is  "responsible  for  the marked point"
and might be considered as a symplectic structure on  the  orbit  of
dressing transformations   -   a   Poisson-Lie  replacement  of
coadjoint action;  the second part of (\ref{sympldouble})  is
related to a "handle" of the torus.  Hence the whole sum
     (\ref{sympldouble}) has the  same  structure  as the symplectic
phase space for Calogero type models  and may be viewed as a
Poisson-Lie "deformation" of it.  To provide  a reduction with this
"non-hamiltonian" momentum (\ref{constraint}) we note that the
analogue of a "free motion"  for  our phase space is given by the
following
     remark.  The Poisson bracket (\ref{lbrackets}) in the terms of
monodromies $A$ and
     $B$ along two circles of the bunch with the prescribed
orientation can
     be rewritten as
\beq
\{ A , B \}= r A \otimes B -
(1 \otimes B)r(A\otimes 1)      +  (A\otimes 1)r_{21}(1\otimes B) +
A\otimes B r
\label{newbrackets}
\eeq
Let us take the traces of the monodromies as the
Hamiltonians $H_{n} = Tr(A^{n})$. We have:
$$
\{ Tr(A^{n}), B \} =
     Tr_{1} \{ A^{n}, B \} = nTr_{1}(A^{n-1} \{ A,B\} )
$$
and obviously
$$
\{ Tr(A^{n}),A \}=0
$$
  The element   $e^{J}$   in (\ref{constraint})   corresponds  to the
exponentiation of the momentum  level for the  Calogero  system
associated
   with $T^{*}G$. If  we  consider  $Tr(A + A^{-1})$  as  a
   Hamiltonian,then tedious but straightforward  computations  show
that
the system at hands is the trigonometric Ruijsenaars' system.

    Now let us  describe  the  duality  property  for  the
finite-dimensional  many-body problems \cite{duality}.
The  particle  systems  under
consideration are CSM  and  Ruijseenars  ones.According
the adopted terminology we will call them as   the nonrelativestic

and
relativistic  systems.As it was shown before for  the  latter

ones  Hamiltonian  is  the
differential operator while for the former it is the difference

one.

Three pairs of the dual systems  were  discussed  in  \cite  {action}
namely nonrelativistic  rational  and  relativistic  trigonometric
which  are  self-dual  ,while  trigonometric  nonrelativistic  and
rational relativistic are dual to each other.The basic equation in
\cite{action} which in a sense provides the dual variables is

\beq \label{momr}
f(\nu)[A,L]=e{\otimes}e-F(A,L)
\eeq

where $ f({\nu})$ is a function of the coupling constant,function  F
and
vector e depend on the particular case.Matrix L is interpreted  as
the  Lax  operator  while  A  is  some  diagonal   matrix.As   usual
the
Hamiltonian is the trace of the power of the Lax operator.

  Transition to the dual system  can be  performed  in  the
following
manner.At first let us diagonalize $L(g,p_{i},q_{i})$

\beq \label{diag}
\widetilde{L}=TLT^{-1}=diag(\lambda_{1},....,\lambda_{N})
\eeq

then using (\ref{momr}) one can found that

\beq \label{alax}
\widetilde{A}=L(-{\nu},{\lambda}_{i},{\widetilde q_{i}})
\eeq

coincides with the Lax operator for another system
with the change of the  sign  of  the  coupling  constant.Now  the
Hamiltonian  of   the   new    system    is    a    function    of
$Tr{\widetilde{A}}$.It  is
important that the procedure above is nothing but the  transition
to the action-angle variables.Indeed it was shown \cite{action} we

have the  linear  flow  in the
new variables

\beq \label{acan}
(\lambda_{i}(t),q_{i}(t))
{\rightarrow}(\lambda_{i},q_{i}+tH^{\prime})
\eeq

   Let us now show that the transition  to  the  dual  system  has  a
natural  interpretation  in  terms  of  Hamiltonian  reduction.The
choice $ T^{*}G$ as a phase space is instructive enough to manifest
the
main   points.The   space $  T^{*}G$   consists   from   the   pairs
$(g{\in}G,A{\in}{\cal G})$
and due to the general prescription let us consider the moment map
in the evident form

\beq \label{momym}
g^{-1}Ag-A={\nu}(1-e{\oplus}e)
\eeq

where the the level of the moment in the r.h.s. is chosen to  get
the Calogero type system we are looking for.One can proceed now in
two different ways.In the first  approach  we  diagonalize  g  and
express A as a function of g from (\ref{momr}).It can be  easily
seen  that for  $g {\in} SU(N)$  and  Hamiltonian $ Tr  A^{2}$   one
has
nonrelativistic trigonometric system.But it is  also  possible  to
diagonalize g and express it as a function of A.Then the choice
$H=Tr(g+g^{-1})$ yields the relativistic rational system.

Now is is almost evident that we have to identify  (\ref{momr})
  and  (\ref{momym})
A simple inspection shows that the first case corresponds to
$A=exp(\mu diag(q_{1},....q_{n}))$  and $ F=0$,while  the
  second  case
results in  $   A=diag(q_{1},......q_{n})$    and $ F=L$
    in
(\ref{momr})
.These are just the functions  used  in  \cite{action}  to  describe
duality between the systems above.Self-dual cases can  be  treated
in a similar manner.

Finally let us mention the change of the coupling constant sign.A

little bit later will give a geometrical interpretation  of  it
but let us remark now that in the quantum case it results in the
change
$g(g-1){\rightarrow}g(g+1)$ in the potential.

Let us turn to consideration of the phase spaces related with
the  affine  groups- $ T^{*}{\cal G}; T^{*}{G} $.The relevant  purely
geometrical
actions are topological YM theory and $G/G$  WZW  correspondingly.So
our goal is to get the dual systems for these patterns.

As before the main tool which provides us with the dual system  is
the corresponding moment map.Recall that for the case of YM theory
it  reads
as

\beq \label{mym}
k{\partial_x}{\phi} +[A,\phi]=\delta(x)J_{\nu}
\eeq

 We take the level of
 the  moment
map which corresponds to the degenerate orbit

\beq \label{level}
J_{\nu}={\nu}{\sum_{\alpha{\in}\Delta_{+}}}(e_{\alpha}+e_{-\alpha})
\eeq

where $ e_{\pm,\alpha}$ are the elements of the nilpotent
subalgebras
of G,which correspond to the roots $ \alpha$ and $ \delta_{+}$ is a
set
of positive roots.Decomposing $\phi$ into  Cartan  D  and  nilpotent
parts $ {\varphi}_{\pm}$ we immediately find ourselves with the
equation for
the nilpotent part

\beq \label{nilpot}
k\partial_x{\varphi_{\alpha}}+<D,\varphi_{\alpha}>=\delta(x)[J_{\nu
}]_{\alpha}
\eeq

where D is diagonal constant matrix.Then after decomposition at $x
\neq 0$

\beq \label{decom}
\phi_{\alpha}(x)=exp(-\frac{x}{k}<D,\alpha>)M_{\alpha}
\eeq

the boundary condition equation appears

\beq \label{bound}
g^{-1}M_{\alpha}g-M_{\alpha}=[J_{\nu}]_{\alpha}
\eeq

which exactly coincides with  the  form  of  the  moment  map  for
$T^{*}G$,where G is the finite-dimensional group.Note that  equation
(\ref{bound}) appears as a consequence of  the  special  form  of
the
level of the moment map.

To get the CSM system  in $ A_{1}$ , "Wilson loop" ,
representation  Hamiltonian is taken in the following form $
H=Tr{\phi^{2}}$.Thus
with (\ref{bound}) as a boundary  condition  in  the  affine
case  it  is  natural   to   look   for   the   system   in  $ \phi$,
"electric field ",
representation.In a sense the only ingredient we need for  is  the
proper choice of the Hamiltonian.The only candidate for this  role
is
 the gauge invariant Wilson loop

\beq \label{WL}
W=Pexp(\int_{S^{1}}dxA_{1}(x))
\eeq

Now we are looking for the solution of (\ref{bound}) with diagonal
M.It
can be shown \cite {gorrui} that the  resolution  of  the  constraint
gives

$$
 \phi_{i}=\frac{1}{N}\prod_{j \neq i}
\frac{(q_{i}-q_{j}-
\frac{\nu}{N})}{ (q_{i}-q_{j})}
$$

moreover  g  coincides  with  the  Lax   operator   for   rational
Ruijsenaars system.Thus the action

\beq \label{action1}
L_{dual}=\int(\phi F+\nu JA_{0}\delta (x) +\xi (W+W^{-1}))
\eeq

results in the rational Ruijsenaars  system for the topological

degrees

of freedom,relevant for the dynamics on the moduli space.It is
evident that if we consider a little bit more general Hamiltonian
with the additional ${\rho}$Tr${\phi}^{2}$ term the corresponding
system

occurs  to
be

\beq
H_{1}=H_{Suth}+\xi (expon.interaction)
\eeq

in "Wilson loop" representation  and

\beq \label{quad}
H_{2}=H_{rat.R-S}+\rho \sum_{i}x_{i}^{2}
\eeq

in  "electric field"    representation.Note    that

(\ref{quad}) is a
finite-difference Schr\"odinger operator.

Let us proceed  now to the case of gauged  WZW  theory.Here  we
have the following constraint

\beq \label{wzwcon}
kg^{-1}\partial g+g^{-1}Ag-A=\delta(x)J
\eeq

Once again the  key  point  is  the  resolution  of  the  boundary
condition equation

\beq \label{resol}
G^{-1}ZGZ^{-1}=exp(\frac {2\pi J}{k})
\eeq

 where $Z=diag(z_{1},...z_{N}),G=G(x=0)$ and
$$
g=exp(xD/k)G(x)exp(-xD/k)
$$

One immediately recognizes in (\ref{resol}) the moment  map  for  the
Heisenberg double.It is also clear that the
corresponding  finite-dimensional  trigonometric

Ruijsenaars system
enjoys the  self-duality  property  which  is  in  agreement  with
(\ref{resol})
.Note also that the selfduality  helps    to  understand  the
origin of  the duality of the zonal spherical functions for quantum
groups
$$
\Psi_{n}(q^{i\nu})=\Psi_{-2i\nu -l}(q^{-\frac{n+l}{2}})
$$
where $ \Psi_{n}$  is  expressed  in  term  of   Rogers-Askey-Ismail
polynomials $C_{n}(cos(\theta;t;q)$ in a following way
$$
\Psi_{n}(e^{i\theta})=\frac{\Phi_{n}(e^{i\theta};t;q)}{\Phi_{n}(t^{
1/2};t;q)}
$$
where
$$
\Phi_{n}(e^{i\theta};t;q)=C_{n}(cos{\theta};t;q)
$$
This means that the finite-difference equation almost  coincides
with
the three-term recurrence  equation  for  the  Macdonald

polynomials.In  our
terms it gets the interpretation as the transition to the dual
representation.More
trivial example of the  same  duality  exists  also  for the rational
Calogero case when the Bessel function has the same property.

 For the GWZW case it  is  instructive  to  have  the  geometrical
interpretation  of  the  duals.It  can  be  achieved   using   the
Chern-Simons interpretation from \cite {gorrui}.For a  CS  theory  on
the torus with a puncture (\ref{resol}) evidently means the monodromy
condition for a A and B cycles and a puncture

\beq \label{monod}
g_{A}g_{B}g_{A}^{-1}g_{B}^{-1}=g_{C}
\eeq

Therefore the duality transformation  is  nothing  but  the
modular transformation for the torus.The change of the  sign $ \nu
{\rightarrow}{-\nu}$  reflects the change of the orientation of the
surface.

 It is amusing that both systems  YM and its dual can  be  derived
from GWZW.Indeed Macdonald polynomials provide the wave  functions

both
for trigonometric Calogero  and  rational  Ruijjsenaars systems

in  the  different
limits;namely $\frac {\nu}{kN}{\rightarrow}{0}$  with $ k=const$
and

$k{\rightarrow}\infty$ $\nu=const$ correspondingly.Note also

that in these
two cases one has dynamics in different configuration

spaces;monodromies
around $  S^{1}$  in "Wilson loop" representation

and the Young tablou  in the "electic field" representation
because eugenvalues of
$\phi$ live  just on this object.The detailed analysis of duality and
its  relation with another problem can be found in \cite{duality}.

\section{Conclusion}

Let us here formulate the main points of our approach once again.One
has to start with some topological field theory with the finite
number degrees of freedom which fix some particular moduli
space.There is one to one correspondence between the hierarchy of the
phase spaces and the ones of the topological theories.It  is
certainly clear that at least all phase spaces in the range from the
cotangent bundle to the finite-dimensional algebra to the double of
the two-loop group have the many-body problem counterpart.It is
important that the actions for these theories are a sum of two
terms,standard pdq term and the moment map("Gauss law") constraint
with some Lagrangian multiplier.Therefore the problem of the
constraint resolution appears which is nontrivial because of the
presence of the sources.Finally the particular many-body system comes
out
if the proper Hamiltonian is added.Coordinates on the phase spaces
appears to be the ones on the moduli spaces of some bundles or
cotangent bundles to the moduli spaces.In all cases above the world
sheet is the torus with one marked point or its degenerations times
the time direction .

The picture is incomplete of course.The question which arises
immediately is related with the generalization to the higher genus
curves with arbitrary number of the marked points.Another challenging
problem is the treatment in the same manner more complicated moduli
space like moduli spaces of instantons and monopoles.
One more important question to be solved is about the proper limiting
procedure to  p-adics.The existence of this limit is implied by the
appearance of Macdonald polynomials in the consideration above.It is
also interesting to include the dynamical fermions into the
problem.In particular some nontrivial phenomena can be expected if
these fermions would be chiral and the anomaly in the Gauss law
constraints appears \cite{anomaly}.The investigation of these
problems is in progress now.

Let us briefly discuss the recent results which are relevant to the
problem under consideration.At first there was a substantial progress
achieved in a purely algebraic approach to the spherical functions on
finite-dimensional and affine algebras and the corresponding Laplace
type operators.In particular it was proved that in all cases the wave
functions of CSM systems can be presented as the trace of the twisted
intertwiner operator between the Verma module and the evaluation
representation \cite {etingof}.Moreover the transparent derivation of
Calogero system from the Bernard equation for the one-point conformal
block in WZW theory on the torus was presented.The close algebraic
analysis with explicit mapping between finite-dimensional problems
and KZ equations can be found in \cite{cherednik}.It demonstrates the
important role of the Hecke algebra and its generalization in the
problem at hands which is still to be clarified.

Another finite-dimensional Haldane-Shastry system also attracts an
attention nowadays
because of its rich symmetry structure \cite{haldane}.It was shown in
\cite {poli2} that such systems with the internal degrees of freedom
appears when the Wilson line is taken in an arbitrary
representation.Symmetries of the theory help in the calculations of
the correlators in Haldane-Shastry  as well as CSM systems

( \cite {pas} an references  therein).Being the system which has the
Lax representation  it also can be present in the r-matrix
formalism.It was done

(\cite {avan} and references therein) and surprisingly the so called
dynamical r-matrix appears in the elliptic case \cite{sklya}.Note
also that the new objects -namely Dunkl and shift operators appear to
be relevant \cite{dunkl}.The interpretation of these operators in
terms of the field theories remains an open question.

The final remark is about the relation between the motion of the
poles in KP system and the classical Calogero dynamics.Since the
pioneer paper \cite{poles}
there was some activity on this subject,which is related with the
deep questions about the moduli spaces (\cite{previato} and
references therein) but a lot of work is still to be done.

I am indebted to V.Fock ,V.Rubtsov and especially to Nikita Nekrasov
for the collaboration.I acknowledge to many people for the
discussions on these subjects,but the conversations with A.Zabrodin,
A.Marshakov and M.Olshanetsky are especially useful.I am grateful

to A.Niemi for the hospitality at the Uppsala University.The work was

partially supported by RFFR grant
and the Soros Foundation grant.

\end{document}